\newcommand{\CLASSINPUTinnersidemargin}{18mm}
\newcommand{\CLASSINPUToutersidemargin}{12mm}
\newcommand{\CLASSINPUTtoptextmargin}{20mm}
\newcommand{\CLASSINPUTbottomtextmargin}{25mm}
\documentclass[10pt,conference,a4paper]{IEEEtran}

\usepackage{times}

\usepackage{graphicx}

\usepackage{amsmath}
\usepackage{times}
\usepackage{graphicx}
\usepackage{amsfonts}
\usepackage{amssymb}
\usepackage{cite}
\usepackage{subfigure}
\usepackage{url}
\usepackage{tikz}
\usetikzlibrary{positioning}
\usetikzlibrary{arrows}

\tikzset{
	crossFran/.pic={
		code={\tikzset{scale=5/10}
			\draw[thick,blue] (0,0) -- (5,5);
			\draw[thick,red] (0,5) -- (5,0);
	}}
}

\tikzset{
    rectangleFran/.pic = {
    code={
        \draw[black](0,0) rectangle(5,2); 
        \path (0.25,0.25)  pic[scale=0.2,color=black] {crossFran};
        \path (0.25,1.25)  pic[scale=0.2,color=black] {crossFran};
        \path (1.25,0.25)  pic[scale=0.2,color=black] {crossFran};
        \path (1.25,1.25)  pic[scale=0.2,color=black] {crossFran};
        \path (2.25,0.25)  pic[scale=0.2,color=black] {crossFran};
        \path (2.25,1.25)  pic[scale=0.2,color=black] {crossFran};
        \path (3.25,0.25)  pic[scale=0.2,color=black] {crossFran};
        \path (3.25,1.25)  pic[scale=0.2,color=black] {crossFran};
        \path (4.25,0.25)  pic[scale=0.2,color=black] {crossFran};
        \path (4.25,1.25)  pic[scale=0.2,color=black] {crossFran};
        }
    }
}

\DeclareGraphicsExtensions{.png,.eps,.ps,.pdf}

\newcommand{\complexnumbers}{\mathbb C}

\begin{document}

\title{Optimizing MIMO Efficiency in 5G through Precoding Matrix Techniques}


\author{
\authorblockN{Francisco D\'iaz-Ruiz, Francisco J. Mart\'in-Vega, Gerardo G\'omez and Mari Carmen Aguayo-Torres}
\authorblockA{Communications and Signal Processing (ComSP) Lab, Telecommunication Research Institute (TELMA), \\ Universidad de M\'alaga, E.T.S. Ingenier\'ia de Telecomunicaci\'on, Bulevar Louis Pasteur 35, 29010 M\'alaga (Spain)}
\authorblockA{ \{fdiaz, fjmvega, ggomez, aguayo\}@ic.uma.es}
}

\maketitle

\begin{abstract}
Multiple-Input Multiple-Output (MIMO) systems play a crucial role in fifth-generation (5G) mobile communications, primarily achieved through the utilization of precoding matrix techniques. This paper presents precoding techniques employing codebooks in downlink MIMO-5G wireless communications, aiming to enhance network performance to meet the overarching 5G objectives of increased capacity and reduced latency. We conduct a comparative analysis of various precoding techniques outlined by the 5G standard through diverse simulations across different scenarios. These simulations enable us to assess the performance of the different precoding techniques, ultimately revealing the strengths and weaknesses inherent in Type I and Type II codebooks.
\end{abstract}


\section{Introduction}

The emergence of the 5G has been pivotal in enabling a wide range of services and applications, characterized by traffic patterns distinct from previous networks. As 5G advances, it not only meets the increasing demands of current applications but also unlocks novel opportunities, reshaping the digital landscape dynamically \cite{5GOpportunities}.

Within 5G, three distinct use cases can be identified depending on how network resources are utilised: enhanced mobile broadband (eMBB), massive machine-type communication (mMTC) and ultra-reliable, low-latency communications (URLCC) \cite{dahlman20205g}. The diverse needs of these categories have driven the development of intricate techniques aimed at improving network capacity and efficiently meeting user demands. This has led to renewed interest and expectations for high-speed, low-latency wireless communication capabilities. One of the key pillars of this technological revolution is the use of MIMO techniques, which significantly improve the spectral efficiency and capacity of wireless networks.

MIMO technology has been widely adopted in the 5G standard to harness spatial multiplexing and improve wireless transmission performance. However, the efficient exploitation of MIMO capabilities in 5G environments presents unique challenges due to the inherent complexity of channel scenarios and the presence of interference among multiple transmitters and receivers.

A key strategy to enhance the efficiency and performance of MIMO in 5G is the use of precoding matrix techniques. These techniques enable the adaptation of the transmitted signal to channel conditions and mitigate interference between antennas, resulting in a significant improvement in received signal quality and system capacity.

In this paper, a detailed performance study of the two types of precoding techniques present in the 5G standard is carried out. During the course of the first section, the system model used to obtain the simulation results is explained, and a comparison is made between the two precoding techniques, showing the throughput obtained with both of them and the required overhead by their reporting in the uplink.




\section{System model}

We are examining a system model based on link-level communication, wherein the transmission of the Physical Downlink Shared Channel (PDSCH) occurs between the Base Station (BS) and a User Equipment (UE) in the downlink, and the UE reports channel information back to the BS.

\begin{figure*}[ht]
\centering
\includegraphics[scale=0.8]{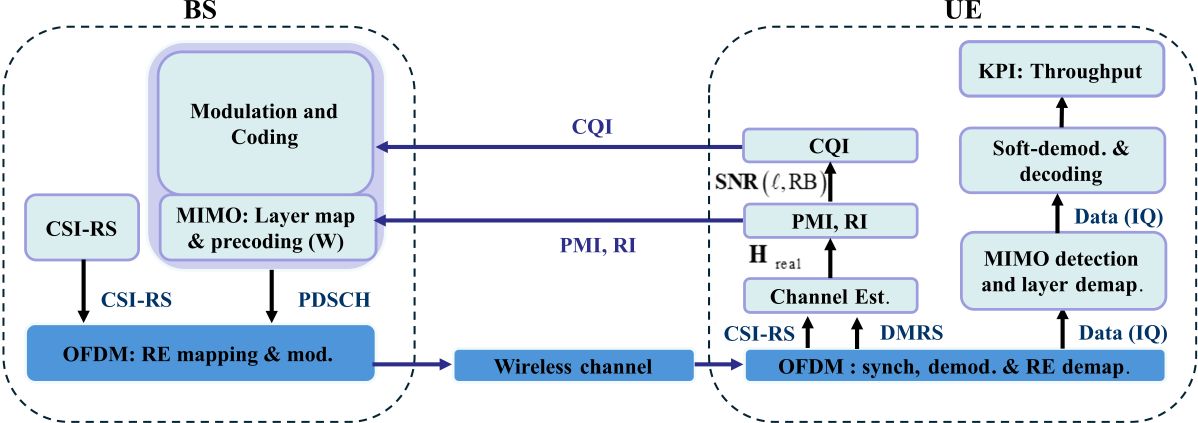}
\caption{Block diagram of the simulator implemented in MATLAB.}
\label{fig:Sketch1}
\end{figure*}

Figure \ref{fig:Sketch1} illustrates the block diagram of the complete system model of a conventional 5G transmission simulator, utilized for dataset generation. The initiation of a transmission entails the crucial selection of modulation and coding rate. Subsequently, after mapping onto the time-frequency resources and employing Orthogonal Frequency Division Multiplexing (OFDM) modulation, the signal traverses the channel. Afterwards, the UE receives and demodulates the signal and estimates the channel by the Channel State Information Reference Signal (CSI-RS).  Finally, the UE computes and reports the following set of indicators: 1) Precoding Matrix Indicator (PMI), which represents an index to a MIMO precoding matrix within the selected codebooks; 2) Rank Indicator (RI), which indicates the suggested number of layers in the next tranmission; and 3) Channel Quality Indicator (CQI), which represents the channel quality.

\label{sec:System Model}

\subsection{MIMO Channel Capacity}

In a MIMO system with $N_{R}$ receiver antennas and $N_{T}$ transmitter antennas the signal received can be expressed in the domain frequency as

\begin{equation}
    \mathbf{Y} = \mathbf{H} \cdot \mathbf{X} + \mathbf{Z},
\end{equation}


\noindent where $\mathbf{Y} \in \complexnumbers^{N_{R} \times 1}$ is the received symbol for each receiver antenna, $\mathbf{X} \in \complexnumbers^{N_{T} \times 1}$ is the transmited symbol for each transmission antenna, $\mathbf{Z} \in \complexnumbers^{N_{R} \times 1}$ is the Additive White Gaussian Noise (AWGN) and $\mathbf{H} \in \complexnumbers^{N_{R} \times N_{T}}$ is the MIMO Channel Matrix.

It will be assumed that transmitter and receiver have perfect knowledge of the channel and that the objective is to transmit the maximum information rate while satisfying a predefined target error rate. As indicated in \cite{cho2010mimo}, the Singular Value Descomposition (SVD) is applied as.

\begin{equation}
    \mathbf{H}=\mathbf{U} \mathbf{\Sigma} \mathbf{V}^{H},
\end{equation}

\noindent where $\mathbf{U} \in \complexnumbers^{N_{R} \times N_{R}}$ y $\mathbf{V} \in \complexnumbers^{N_{T} \times N_{T}}$ are unitary matrix and $\mathbf{\Sigma} \in \complexnumbers^{N_{R} \times N_{T}}$ is a diagonal matrix whose elements are the singular values of matrix $\mathbf{H}$. Since $\rho$ is the rank of the channel matrix ($\rho = \mathrm{Rank}(H) \leq \mathrm{min(N_{R},N_{T})}$), then there are $\rho$ singular values.

The transmitter applies the matrix V as precoding matrix whereas the receiver uses U as detection matrix. The capacity of the MIMO channel assuming symbols with unity power can be expressed as

\begin{equation}
        C = \sum_{i=1}^{\rho} \mathrm{log}_2 \left( 1 + \frac{\sigma_{i}^{2}}{\sigma_{n}^{2}} \right) \,\, \mathrm{bps/Hz},
\end{equation}

The equivalent of a MIMO channel after precoding and ideal detection resembles that of $\rho$ independent channels, enabling the transmission of symbols. Each independent channel for information transmission is termed a layer and its gain is denoted by $\sigma_{i}^{2}$. The term $\sigma_{n}^{2}$ refers to noise. Depending on the number of layers utilized, the transmission is categorized as follows:

\begin{itemize}
    \item \textbf{Spatial multiplexing}: when more than one layer is utilized for transmission.
    \item \textbf{Beamforming}: when transmission is restricted to a single layer usage.
\end{itemize}



\subsection{Codebooks in 5G}
In the context of reference signal reporting, a codebook is a set of precoding matrices that are applied at the start of symbol sequence transmission in order to maximise channel capacity. The optimal precoding matrix, as seen in the previous section, would be the one obtained from the Singular Value Decomposition (SVD) process, but then an infinite capacity CSI report would be required. Since this is not possible, the standard defines sets of precoding matrices that are defined in the codebooks to facilitate reporting. The reporting of the precoding matrix is done by means of the PMI, which is a set of indices that define the precoding matrix within the codebook being used. The codebook depends on the spatial structure of the antenna array (number of antennas and shape of the panels), number of layers and codebook type. In 5G there are two types of downlink codebooks:

\begin{itemize}
    \item \textbf{Type I}: provides acceptable spatial resolution with low reporting overheads.
    \item \textbf{Type II}: offers a higher spatial resolution than type I, but at the same time increases the reporting burden.
\end{itemize}

\begin{figure}
    \centering
    \begin{tikzpicture}[node distance = 7cm, auto, place/.style={circle,draw=black,fill=black,thick,radius=2pt}, scale=0.70, every node/.style={transform shape}]
        \path (0,0) pic[color=black] {rectangleFran};
        \path (5.25,0) pic[color=black] {rectangleFran};
        \draw[<->, black](0,2.25) -- node [text width = 2cm, align=center, above] {\Large{$N_1=5$}} (5,2.25);
        \draw[<->, black](-0.25,0) -- node [text width = 2cm, align=center, left, xshift=0.20cm] {\Large{$N_2=2$}} (-0.25,2);
        \draw[<->, black](0,-0.25) -- node [text width = 2cm, align=center, below] {\Large{$N_g=2$}} (10.25,-0.25);
    \end{tikzpicture}
    \caption{Parameters for the selection of different codebook configurations as a function of the cross-polarised antenna array structure at the BS.}
    \label{fig:ConfigurationAntennaParmeters}
\end{figure}
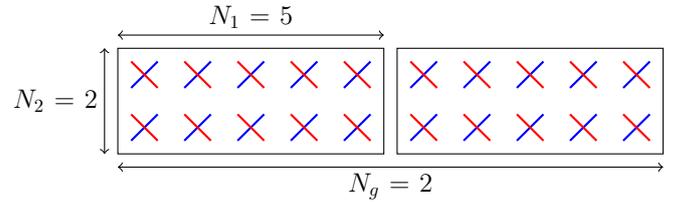

The selection of the codebook is therefore based on the selection of the codebook type, on the antenna structure $N_1$, $N_2$, $N_g$, where $N_1$ is the number of antennas horizontally, $N_2$ the number of antennas vertically and $N_g$ the number of antenna panels (represented in Fig. \ref{fig:ConfigurationAntennaParmeters}.) and on the number of layers, which is defined by the L.

\begin{equation}
    \mathcal{W}(C,N_{1},N_{2},N_{g},L) = \left\{ \textbf{W}_{1},...,\textbf{W}_{M} \right\},
\end{equation}
where $\mathcal{W}$ is the set of pre-coding matrices available in the selected codebook.

The PMI indices uniquely identify a single codebook matrix, where the indices $i_1$ refer to the precoding matrix for the whole bandwidth and the indices $i_2$ refer to the precoding matrix for each sub-band, whereby the matrix used for precoding is the result of the mutiplication of both matrices:

\begin{equation}
    \mathbf{W}_{i(\epsilon)} = \mathbf{W}_{i_{1}} \mathbf{W}_{i_{2}(\epsilon)}
\end{equation}

For the implementation in the simulator of the PMI and RI report calculation, MATLAB's 5G Toolbox \cite{5GToolbox} was used, which is based on an exhaustive search among the possible RIs and PMIs for the current channel estimation and choosing the configuration that maximises the SNR and Block Error Rate (BLER) to stay below a target BLER.

\subsection{Overhead in 5G report for precoding matrix}

The calculation of precoding matrix reporting overhead is contingent upon several factors, including the type of report utilized, the number of transmission layers determined by the RI, and the antenna configuration. The overhead calculation entails interpreting the various indices that need to be reported based on the PMI report configuration. This complexity arises from the challenges associated with precoding matrix reporting and its interpretation within the standard. The overhead is derived through the following steps:

\begin{enumerate}
    \item Identify the oversampling factor reflect in 3GPP standard in function of antennas configuration. This factor oversampling is shown in Table 5.2.2.2.1-2 in \cite{3gpp_38214}.
    
    \item A worst-case estimate is made of the number of overhead bits in the report based on the configuration used in simulation. The report depends on the configuration that has been used in simulation. The number of overhead bits depends the number that represent each index at the PMI report, obtained from \cite{3gpp_38214}.
  
    \item Calculate the average overhead caused by PMI report, by the next equation:
    \begin{equation}
        \mathbb{E}[Q] = \sum_{i=1}^{L} p_{i} \cdot Q_{i},
    \end{equation}
    where $\mathbb{E}$ is the expectation operator and $Q$ is the mean overhead caused by PMI report, $Q_{i}$ is the number of overhead bits for each layer and $p_{i}$ is the probability of selecting $i$ layers.

\end{enumerate}
\begin{itemize}
\item \textbf{Type I}: The number of bits used for PMI reporting in type I can be obtained from: 
    \begin{equation}
    \label{eq:OverheadIndicesTypeI}
    \begin{aligned}
        i_{1,1} &= \mathrm{log}_{2}(N_{1}O_{1}) \\
        i_{1,2} &= \mathrm{log}_{2}(N_{2}O_{2}) \\
        i_{1,3} &= 0 & v = 1 \\
        i_{1,3} &= \mathrm{log}_{2}(4)  & v \in \left\{ 2,3,4 \right\} \\ 
        i_{2}   &= \mathrm{log}_{2}(4)  & v = 1 \\
        i_{2}   &= \mathrm{log}_{2}(2)  & v \in \left\{ 2,3,4 \right\}
    \end{aligned}
    \end{equation}
\item \textbf{Type II}: The main difference between Type I and Type II is that Type I selects a specific beam from a group of beams and Type II selects a group of beams and makes a linear combination of the selected beams. This is why for Type II a report containing more information is required. The overhead caused by Type II can be calculated as:
    \begin{equation}
    \label{eq:OverheadIndicesTypeII}
    \begin{aligned}
        i_{1,1} &= \mathrm{log}_{2}(O_{1}) + \mathrm{log}_{2}(O_{2}) \\
        i_{1,2} &= \mathrm{log}_{2}\binom{N_{1}N_{2}}{B} \\
        i_{1,3,1} &= \mathrm{log}_{2}(B)  &if \quad  L = 1\\
        i_{1,4,1} &= 2B\mathrm{log}_{2}(8) &if \quad  L = 1\\
        i_{1,3,2} &= \mathrm{log}_{2}(B)  &if \quad  L = 2\\
        i_{1,4,2} &= 2B\mathrm{log}_{2}(8) &if \quad  L = 2\\
        i_{2,1,1} &= 2B\mathrm{log}_{2}(N_{PSK})  &if \quad  L = 1\\ 
        i_{2,2,1} &= 2B\mathrm{log}_{2}(2)  &if \quad  L = 1\\ 
        i_{2,1,2} &= 2B\mathrm{log}_{2}(N_{PSK})  &if \quad L = 2 \\ 
        i_{2,2,2} &= 2B\mathrm{log}_{2}(2) &if \quad  L = 2\\ 
    \end{aligned}
    \end{equation}
    
    In equations (\ref{eq:OverheadIndicesTypeI}) and (\ref{eq:OverheadIndicesTypeII}), $B$ is the number of beams to be reported, $L$ is the number of layers indicated by RI and $O_1, O_2$ is the oversampling factor.
\end{itemize}

\section{Simulation results}

During the course of this section, the aim is to test the difference between the selection of the pre-coding matrix using different types of codebooks. The objective is to evaluate the improvement caused by the use of Type II as the spatial diversity increases due to the increase of antennas in reception and transmission, due to the higher spatial resolution that this type of codebook allows. The Type II codebook presents a better spatial resolution increasing the overhead of the report, this spatial resolution makes that in systems with a great spatial diversity, that is to say, a high number of antennas in transmission, improves the conditions compared to Type I. For the simulation, the configuration of the parameters is presented in Table \ref{tab:SimulationParameters}. To obtain these graphs, a signal to noise ratio sweep was performed and the average throughput of the different transmissions was obtained, comparing the performance of the different configurations.

\begin{table}[ht]
\renewcommand{\arraystretch}{1.5}
\caption{Simulation parameters.}
\label{tab:SimulationParameters}
\centering
    \begin{tabular}[width=\columnwidth]{ l l }
    \textbf{Parameters} & \textbf{Values (units)}\\
    \hline
     Multipath Channel Model & CDL-A \cite{3gpp38901}\\
     \hline
     Simulation duration & 10000 (slots) \\
     \hline
     Maximum Doppler Shift & 5 (Hz) \\
     \hline
     Delay Spread & 100 (ns) \\
     \hline
     Number of transmitting antennas & 8 \\
     \hline
     Number of receiving antennas & [2,8] \\
     \hline
     Number of layer in tranmission ($L$) & [1-8] \\
     \hline
    \end{tabular}
\end{table}

\begin{figure}[h]
    \centering
    \includegraphics[width=0.95\columnwidth]{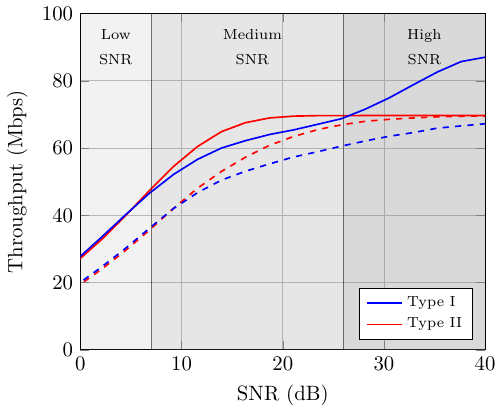}
    \caption{Comparison of Type I codebook versus Type II codebook in MIMO scenarios, in solid line 8x4 and in dashed line 8x2 scenario.}
    \label{fig:ComparisonCodebook}
\end{figure}

\begin{figure}[h]
    \centering
    \includegraphics[width=\columnwidth]{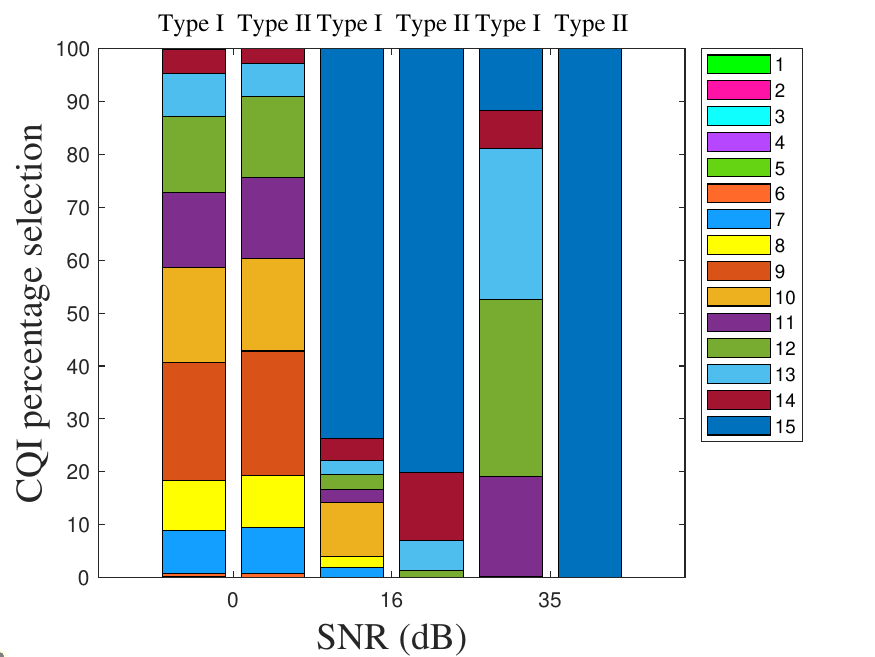}
    \caption{Comparison of the percentage probability of CQI selection for the two types of coding at three different points in each of the regions of Fig. \ref{fig:ComparisonCodebook}, first column is for Type I and second is for Type II. The different colors represent the possible CQIs.}
    \label{fig:ComparisonProbabilityRI}
\end{figure}

\begin{figure}[h]
    \centering
    \includegraphics[width=\columnwidth]{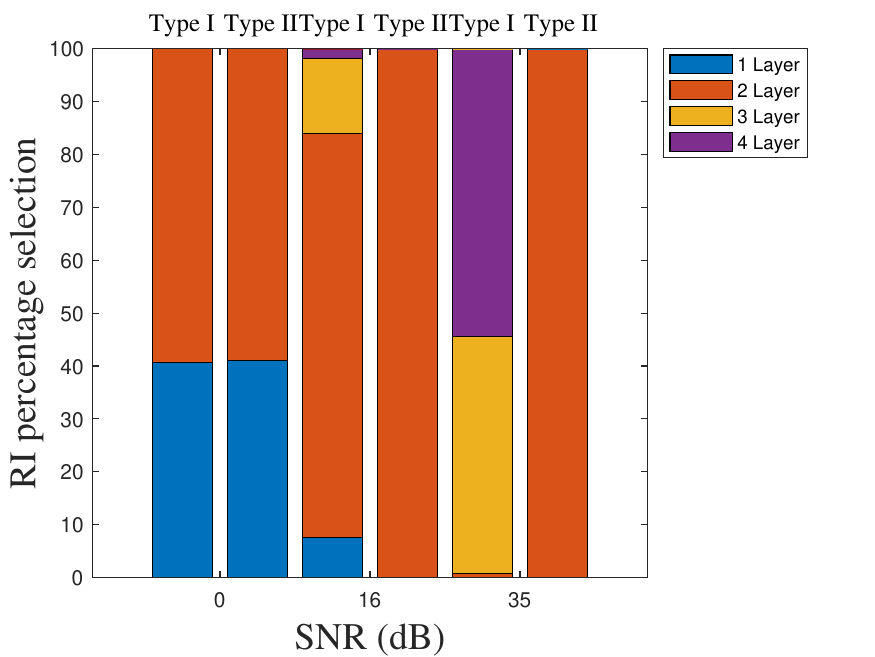}
    \caption{Comparison of the percentage probability of RI selection for the two types of coding at three different points in each of the regions of Fig. \ref{fig:ComparisonCodebook}, first column is for Type I and second is for Type II.}
    \label{fig:ComparisonProbabilityCQI}
\end{figure}

Fig. \ref{fig:ComparisonCodebook} illustrates the comparison of the throughput obtained from Type I versus Type II precoding matrices. Additionally, this evaluation is conducted across two distinct MIMO scenarios, with the dashed line representing an 8x2 scenario and the solid line depicting an 8x4 scenario and Fig. \ref{fig:ComparisonProbabilityCQI}, Fig. \ref{fig:ComparisonProbabilityRI} show the comparison of the percentage of RI and CQI selection at different SNR points for different codebook types in the 8x4 scenario, which justifies the obtained throughput results.
 
In the 8x4 scenario depicted in Fig. \ref{fig:ComparisonCodebook}, three distinct regions are identified, each demonstrating changes in system performance concerning throughput:

\begin{itemize}
\item \textbf{Low SNR region}: Below 7 dB, the throughput achieved with different precoding techniques remains similar. This is attributed to the degradation of channel estimation performed on the pilots by the noise, diminishing the spatial resolution gain of Type II.
\item \textbf{Medium SNR region}: Between 7 dB and 26 dB, Type II demonstrates an enhancement in throughput compared to Type I. This improvement primarily stems from the spatial resolution enhancement achieved by Type II. As illustrated in Fig. \ref{fig:ComparisonProbabilityRI} and Fig. \ref{fig:ComparisonProbabilityCQI}, at the 16 dB point, Type II selects higher CQIs, leading to the selection of higher modulation levels and lower coding rates, with all transmissions utilizing 2 layers.
\item \textbf{High SNR region}: Above 26 dB, the throughput of Type I surpasses that of Type II. This is predominantly due to the restriction imposed by the standard on Type II, limiting it to a maximum of two layers. As shown in Fig. \ref{fig:ComparisonProbabilityRI} and Fig. \ref{fig:ComparisonProbabilityCQI}, Type I utilizes from three to four layers while Type II is constrained to two layers, as dictated by the standard.
\end{itemize}

\begin{figure}[h]
    \centering
    \includegraphics[width=0.95\columnwidth]{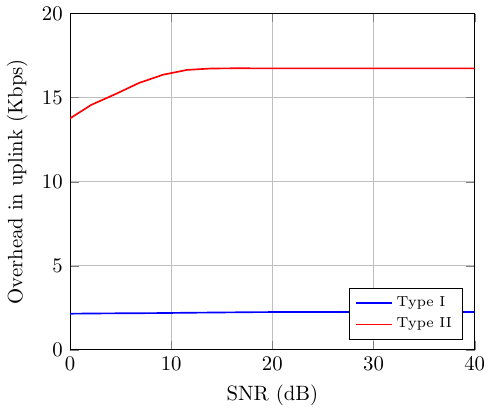}
    \caption{Comparison of the overhead caused by the type of reporting in the uplink.}
    \label{fig:ComparaisonOverheadCodebook}
\end{figure}

Fig. \ref{fig:ComparaisonOverheadCodebook} shows the weakness of Type II compared to Type I, as it requires a higher number of indices to report, resulting in higher uplink overhead.

\section{Conclusions}

Comprehending the various report types associated with the precoding matrix and deriving an overhead calculation for them demonstrates the effectiveness of Type II reporting in enhancing spatial resolution for users. This implies that under circumstances with a favorable average signal-to-noise ratio, the system capacity improves in comparison to Type I reporting, albeit at the cost of increased overhead. Thus, this highlights a trade-off between the overhead incurred by uplink reporting and the enhancement in uplink capacity.

\section*{Acknowledgements}
This work has been supported by Keysight Technologies and by MCIN/AEI/10.13039/501100011033, FEDER \textit{A way of making Europe}, Junta de Andaluc\'ia and University of M\'alaga (UMA) through grants PID2022-137522OB-I00, RYC2021-034620-I and MA/INV/0007/2022.

\bibliographystyle{IEEEtran}
\bibliography{IEEEabrv,biblio}

\end{document}